\documentclass[copyright,creativecommons]{eptcs}
\providecommand{\event}{CCA 2010}
\usepackage{t1enc,color,array,rotating,graphicx}
\usepackage{latexsym,amssymb,amsmath,amsfonts,stmaryrd}
\usepackage[utf8]{inputenc}
\usepackage[english]{babel}

\numberwithin{equation}{section}

\ifx\pdfoutput\undefined\else\ifnum\pdfoutput=1
\fi\fi

\def\newblock{\hskip .11em plus .33em minus .07em}

\newcommand{\Nat}{\mathbb{N}}

\newcommand{\Real}{\mathbb{R}}

\newcommand{\Cplx}{\mathbb{C}}

\newcommand{\GLA}{\begin{eqnarray*}}
\newcommand{\GLE}{\end{eqnarray*}}

\title{Making big steps in trajectories}

\author{
Norbert Th.\ Müller\thanks{This research was partially supported by the 
DFG projects  446 CHV 113/240/0-1 and 445 SUA 113/20/0-1}
  \institute{
    Abteilung Informatik, FB IV\\
    Universität  Trier, Germany
\email{mueller@uni-trier.de}
}
\and 
Margarita Korovina\thanks{
  This research was  partially supported  by  EPSRC grant EP/E050441/1,
  DFG-RFBR
(grant No 436 RUS 113/1002/01, grant No 09-01-91334), RFBR grants  
07-01-00543, 08-01-00336.
  }
  \institute{
   CICADA\\
   University Manchester, UK
   \email{\begin{tabular}{c}Margarita.Korovina@manchester.ac.uk\end{tabular}} }
}
\def\titlerunning{Making big steps in trajectories}
\def\authorrunning{N.Th. Müller \&  M. Korovina}
\begin{document}
\maketitle

\begin{abstract}
We consider the solution of initial value problems within the context of hybrid systems
and emphasise the use of  high precision approximations  (in software for exact real arithmetic).
We propose a novel algorithm for the computation of trajectories up to the area where
discontinuous jumps appear, applicable for holomorphic flow functions.
Examples with a prototypical implementation illustrate that the algorithm might provide results with 
higher precision than well-known ODE solvers at a similar computation time.
\end{abstract}

\section{Introduction}
The central idea underlying  hybrid systems is that of a system of differential equations
(initial value problems, IVP) 
enhanced with the ability to do discontinuous jumps, simular to finite automata. 
Unfortunately, from the viewpoint of 
TTE ({\em e.g.} \cite{BHW07}) discontinuity implies non-computability, which has been investigated 
in detail  in \cite{Collins-2008a}, {\em e.g.}.
Nevertheless, the importance of these systems
forces us to deal with them and provide the best solutions possible.

There do exist many software tools for hybrid systems 
(see \cite{LL09} {\em e.g.});
however, almost all of them are based on {\small\verb#double precision#} numbers; 
a notable exception is \verb#Ariadne# \cite{BCC06} using generic programming, 
it is hence prepared for other implementations of real numbers.

One basic aspect of the hybrid systems is their evolution in time, {\em i.e.} the computation 
of trajectories. In this paper we present an algorithm (implemented using the iRRAM package) 
for {\em efficient and arbitrarily precise 
solutions} of the underlying IVPs up to the area where discontinuous
jumps appear. There are two reasons for using high precision solutions: firstly,
low precision might lead to incorrect assumptions about the location of these jumps points; and secondly -- perhaps unexpectedly -- high intermediate precision can sometimes increase the efficiency.

To illustrate the second aspect, consider the well-known Runge-Kutta methods used for the solution of IVPs. These methods are of fourth order, {\em i.e.}, the error depends on a bound on higher derivatives of the solution as well as on the fourth power of the step width.
Although they are a reasonable choice if applied to
{\small\verb#double precision#} numbers, they will not always be optimal for
higher precision solutions. 
As the step width has to be chosen according to the desired precision of the solution, 
methods with a fixed order lead to the number of steps growing exponentially in the number of bits of the solution.
If the order can be chosen dynamically and arbitrarily high,
significantly fewer steps associated with a much bigger step width are possible, which can lead to a 
polynomial time complexity \cite{MM93}.

Differential equations have been addressed numerous times in computable analysis, for example 
see \cite{BHW07,We00}, where general questions of computability are addressed. 
A very important related result can be found in \cite{Ko83}: 
The computation of solutions of differential equations is closely related to the problem `\#P=FP' from discrete complexity theory.
This immediately implies that for general IVPs we cannot expect to find algorithms that run  in polynomial time.
For special IVPs, on the other side, it is well-known that the solutions are computable in polynomial time. A very detailed
analysis of the resulting complexity for one-dimensional solutions can be found in \cite{MM93}:
If the flow function of the IVP is holomorphic and computable in polynomial time, then we are able to use
methods with arbitrarily high order to solve the IVP and get a polynomial time solution.

In this paper we will take the result from \cite{MM93} and generalise it to IVPs of arbitrary finite dimension. This generalization then is used as the fundamental part of a new algorithm for the computation of trajectories in hybrid systems.
In addition to a discussion of the theory behind our approach we actually present a prototypical 
implementation in the iRRAM software package \cite{Mu00,Mue09}. As an example, we use a rather simple linear differential equation 
where even an analytical solution is known. This allows us to compare our implementation
with conventional IVP solvers, where our prototype has already shown unexpected efficiency 
at an always superior precision.

Implementations of IVP solvers in exact real arithmetic, giving arbitrary precision results, are very rare.
A prototypical implementation mentioned  in \cite{EP07} can hardly be useful in practice, as it seems
to be based on the explicit construction of the solution using piecewise linear functions. 
This will necessarily lead to a complexity that is exponential in the precision of the solution.
Vaguely similar approaches shown in the tutorial section during the CCA 2009 conference in Ljubljana 
were already unable
to compute more than 4 decimals of the integral $\int_0^1f(t)dt$ for the simple function $f(t)=t^2$.
This leads us to assume that in this paper we actually present the first usable implementation for IVPs using exact real arithmetic.
We should mention here that IVP solvers using interval arithmetic (hence also correct, but not arbitrarily precise)
are well-known, for an overview see \cite{NJC99}, {\em e.g.}.

\section{Hybrid Systems}
A hybrid system can be defined as a tuple $H = (Q, {\bf X}, {\bf D}, {\bf G}, {\bf F}, {\bf R})$ 
consisting of  a finite index set $Q$,
a  continuous state space ${\bf X} = \bigcup_{q\in Q} X_q$,  
a collection of invariant domains ${\bf D}=\{D_q\}_{q\in Q}$, $D_q\subseteq X_q$,
a collection of guard sets ${\bf G}=\{G_q\}_{q\in Q}$,  $G_q\subseteq X_q$,
a collection of continuous dynamics or flow conditions ${\bf F}=\{F_q\}_{q\in Q}$ 
defining differential equations in ${\bf X}$,
and a collection of discrete dynamics or reset relations ${\bf R}=\{ R_{q,q'\in Q}\}$, $R_{q,q'}\subseteq 
X_q\times X_{q'}$, 
see e.g. \cite{SS99}.
The part that we are addressing in this paper are the flow conditions  ${F_q}$ that lead to trajectories
in a component $X_q$ of the state space. Whenever such a trajectory enters the guard set $G_q$, 
a discontinuous jump according to ${\bf R}$ 
might happen, quite similar to a non-deterministic automaton with state space $Q$.

In the following, we will consider single trajectories of such hybrid systems. Our goal is not to discuss their 
computability or to formally consider their (theoretical) 
computational complexity, but we want to get a usable 
implementation that computes such trajectories within a component $X_q$ until they enter the guard set $G_q$.
As we do not attempt to use the reset relations, we will not really need $Q$, and  in consequence we will
omit all references to $Q$ in the following.  

Below we will shortly describe the
data structures we use; they will implicitly impose restrictions on the 
hybrid systems we can address:
\begin{itemize}
\item State space ${X}$: 
We will use $X= \Real\times \Real^d$ for an integer dimension $d\in\Nat$. 
In the implementation, we use (dynamically sized) vectors of real numbers.
$d$ is then not given explicitly, but can be derived from the size of the vectors. 
The first component of $X$ will always be used as the time parameter. 
\item Invariant domain $D$: For the flow conditions $F$ that we are able to address at the moment, it would be very artificial to use a proper subset of $X$ here. For simplicity, we only use $D=X$.  

In the following we will nevertheless mention where 
extensions for a non-trivial $D$ would be necessary. In general,  
the distance function $d_{X\setminus D}(\xi)$ from vectors $\xi\in X$ to the exterior of $D$ should be  
computable, implying that $D$ should be  computably open.

\item Flow conditions $F$: The basic property of the flow conditions is that there are  $d$
functions $F_\nu: {\subseteq} (\Real\times\Real^d)\to\Real$ defining differential equations.
We will assume that the $F_\nu$ are holomorphic, hence they can be smoothly extended to complex arguments. 
In the cases we are able to deal with now, 
we will even use that the $F_\nu$ are holomorhpic on the whole set $\Cplx^{d+1}$. 

If the $F_\nu$ are only holomorphic in a restricted area $D_\Cplx$ (with $D\subseteq D_\Cplx$), the distance function $d_{\Cplx^{d+1}\setminus D_\Cplx}(\xi)$ from vectors $\xi\in D$ to the exterior of $D_\Cplx$ should be computable.
 
At the moment, our implementation is restricted to flow functions $F_\nu$ 
that are in fact polynomials (in $d{+}1$ variables and with computable coefficients). 
We will use the following data to get access to the relevant properties of $F$:
\begin{itemize}
\item Maximal degree $\mu$ of the polynomials
\item Single coefficients of the polynomials, {\em i.e.}, we have direct access to the $d\cdot(\mu+1)^{d+1}$ real numbers $c_{\nu,k,i_1,\ldots,i_d}$ for 
$1\leq \nu\leq d$ and $0\leq i_1,...,i_d,k\leq \mu$. 
In applications, 
many of these numbers will be known to be zero. Our later examples of linear homogeneous IVPs will even be restricted to
only $d\cdot(d+1)$ values that may be non-zero.
\item A function $U_F$ on states $(t_0,\overline w_0)\in X$ and distances $\delta,\epsilon\in\Real^+$ 
returning an upper bound for the flow  functions $F_\nu$ on the compact set  
$$C((t_0,\overline w_0),\delta,\epsilon):= \{(t,\overline z)\in\Cplx^{d+1}:
 |t-t_0|\leq \delta \wedge|\overline z - \overline w_0 |\leq \epsilon\}$$
Such a function can obviously be computed directly using the coefficients, implying that $U_F$ would be 
superfluous in a minimalistic setting. 
Nevertheless, we will later see that $U_F$ plays a special role in the solutions 
and should also be helpful in cases that are not yet covered by our implementation.

\end{itemize}
\item Guard set $G\subset X$: $G$ will be given by  an algorithm to compute  $d_G(\xi)$, which implies that 
$G$ is restricted to computably closed sets. Additionally, for a part of our algorithm we will use 
that its interior $G^o$ is computably open; so also the distance $d_{X\setminus G^o}(\xi)$ to the complement of
$G^o$ must be computable.

\item Trajectories: Given an initial state $\xi=(t_0,\overline{w_0})$, our goal is to
to determine how long the trajectory $\overline y$ through $(t_0,\overline{w_0})$ lives until it enters the guard set $G$; we want to find the first $t>t_0$ such that $\overline y (t)$ is in $G$. 
Such a trajectory   $\overline y$ is a vector $(y_1,\ldots,y_d)$ of (possibly partial) real valued functions 
$y_\nu: {\subseteq}\Real\to \Real$ on a real variable $t$ that is usually interpreted as a `time' parameter.
As $\overline y$ touches  $(t_0,\overline{w_0})$, $\overline y$  is (a part of) the solution
of the IVP with flow conditions $F$ and initial condition  $\overline y(t_0)=\overline{w_0}$.
In our setting, it will be natural to extend this to complex variables $t$ and to consider $y_\nu: {\subseteq}\Cplx\to\Cplx$
instead.
\end{itemize}

\section{Solving IVPs}

The basis of the approach we take is a well-known recursion that can be found in mathematical textbooks like \cite{BS85}.
In \cite{MM93} we used it to derive an algorithm showing polynomial time complexity for certain one-dimensional IVPs.
We now generalise this result for higher dimensional systems. The polynomial complexity could also
be provable in these cases, but in this paper we concentrate on an actual implementation.

\subsection{General solution for holomorphic flow functions}

Suppose we have $d$ functions $F_\nu:\Real\times\Real^d\to \Real$ 
($1{\leq}\nu{\leq} d$) and a vector $\overline {w}_0=(w_{1,0},\ldots,w_{d,0})$ such that our IVP has the form 
\begin{equation}\label{IVP} 
\dot y_\nu(t) = F_\nu(t,y_1(t),\ldots y_d(t))~~~~,~~~~ y_\nu(0)=w_{\nu,0} ~~~~~(1{\leq}\nu{\leq} d).
\end{equation}

Note carefully that here we consider the special case where the initial condition $(t_0,\overline {w}_0)$
is restricted to $t_0=0$. The general case of specifying an arbitrary time $t_0$ 
will be addressed later.

The Picard–Lindelöf theorem guarantees a unique solution on some interval containing $t_0=0$ 
if the function vector $F$ and its partial derivatives are continuous on a region around $(t_0,\overline {w_0})$. 
This theorem can be used to get an iterative algorithm for the construction of solutions, 
like in \cite{EP07}.
We will now use much stronger conditions: 
Our assumption is that the functions $F_\nu$ are holomorphic,
{\em i.e.}, there exists a system of coefficients $c_{\nu,k,i_1,...,i_d}$ such that 
for $t\in\Real$ near the origin, for $\overline x=(x_1,\ldots x_d)\in\Real^d$ and for each coordinate $\nu$ ($1{\leq}\nu{\leq} d$) we have 
\begin{equation}\label{IVP0} F_\nu(t,\overline x)=\sum_{k,i_1,...,i_d\in\Nat}
\left[c_{\nu,k,i_1,...,i_d}\cdot t^k\cdot x_1^{i_1} \cdot\ldots\cdot x_d^{i_d}\right].
\end{equation}
We assume that the circle of convergence is large enough to contain the initial condition $(0,\overline {w_0})$.
(Centering the circle of convergence at $\overline {w_0}$ as well as using an arbitrary $t_0$ will be addressed later.)
Then 
\begin{equation} \dot y_\nu(t) =  \sum_{k,i_1,...,i_d\in\Nat}\left[ c_{\nu,k,i_1,...,i_d}\cdot t^k\cdot 
(y_1(t))^{i_1}\cdot\ldots\cdot (y_d(t))^{i_d}\right].\label{eq1}
\end{equation}
As the $F_\nu$ are holomorphic, the functions $y_\nu(t)$ are also holomorphic. 
This follows from `local' versions of the Picard-Lindelöf theorem in the complex plane, 
but we can also just use a Taylor series approach and prove that the radius of convergence is not zero
(which is essentially done in section 4 below). 
So, let $a_{\nu,n}$ be the  corresponding sequences of coefficients, such that
\begin{equation}
 y_\nu(t)=\sum_{n\in\Nat}a_{\nu,n}\cdot t^n.
\end{equation}
The coefficients $a_{\nu,0}$ are already known from the initial condition, as we have
\begin{equation}
\label{initial condition}
a_{\nu,0}=y_{\nu}(0)=w_{\nu,0} ~.
\end{equation}
In the following, we address the coefficients $ a_{\nu,n}$ with $n>0$.
From equation (\ref{eq1}) we see that we need the powers $(y_\nu(t))^i$. 
The corresponding coefficients will be denoted by $a^{(i)}_{\nu,n}$, so 
\begin{equation}
 (y_\nu(t))^{i}=\sum_{n\in\Nat}a^{(i)}_{\nu,n}\cdot t^n ~.
\end{equation}
Comparing coefficients, we get the following recursion for the coefficients of the powers:
\begin{equation}\label{coeff-recursion}
\begin{array}{lcl}
a^{(0)}_{\nu,n} &=& n\mbox{-th value of the sequence }1, 0,0,0, \ldots ~,\\
a^{(i+1)}_{\nu,n} &=& \displaystyle\sum_{0\leq j\leq n} a_{\nu, j}\cdot a^{(i)}_{\nu,n-j} ~.
\end{array}
\end{equation}
It is worth noting that each $a^{(i)}_{\nu,n}$ is determined by the  values
$a_{\nu,j}$ with $j\leq n$. 
A reformulation of (\ref{eq1}) now leads to:
\GLA \dot y_\nu(t) &= & \sum_{k,i_1,...,i_d\in\Nat}\left[ c_{\nu,k,i_1,...,i_d} \cdot t^k\cdot
(y_1(t))^{i_1}\cdot\ldots\cdot (y_d(t))^{i_d}\right]\\
&=&
\sum_{k,i_1,...,i_d\in\Nat}
\left[c_{\nu,k,i_1,...,i_d}\cdot t^k\cdot
\left(\sum_{n_1}a^{(i_1)}_{1,n_1}\cdot t^{n_1}\right)
\cdot\ldots\cdot 
\left(\sum_{n_d}a^{(i_d)}_{d,n_d}\cdot t^{n_d}\right)
\right]\\
&=&
\sum_{k,i_1,...,i_d\in\Nat}
\sum_{\ell\in\Nat}
\sum_{\substack{n_1,n_2,...,n_d\in\Nat\\ n_1+...+n_d+k=\ell}}
\left[c_{\nu,k,i_1,...,i_d}\cdot t^k\cdot a^{(i_1)}_{1,n_1}\cdot t^{n_1}\cdot\ldots\cdot a^{(i_d)}_{d,n_d}\cdot t^{n_d}\right]\\
&=&
\sum_{\ell\in\Nat} t^\ell\cdot 
\sum_{\substack{k,n_1,n_2,...,n_d\in\Nat\\n_1+...+n_d+k=\ell}}
\sum_{i_1,...,i_d\in\Nat}
 \left[c_{\nu,k,i_1,...,i_d}\cdot a^{(i_1)}_{1,n_1}\cdot\ldots\cdot a^{(i_d)}_{d,n_d}\right] ~.
\GLE
Comparing coefficients with $\dot y_\nu(t)={\displaystyle\sum_{\ell\in\Nat}} (\ell{+}1)\cdot a_{\nu,\ell+1}\cdot t^{\ell}$ we see that
\begin{equation}
\label{compare_coeff}
a_{\nu,\ell+1}=\frac{1}{\ell+1} 
\sum_{\substack{k,n_1,n_2,...,n_d\in\Nat\\ n_1+...+n_d+k=\ell}}~~
\sum_{i_1,...,i_d\in\Nat}
 \left[c_{\nu,k,i_1,...,i_d}\cdot a^{(i_1)}_{1,n_1}\cdot\ldots\cdot a^{(i_d)}_{d,n_d}\right] ~.
\end{equation}
Remembering that we already have $a_{\nu,0}=w_{\nu,0}$ and that the values $a^{(i_j)}_{j,n_j}$
needed for $\ell+1$  only depend on
$a_{j,\mu}$ for $\mu\leq n_j(\leq\ell)$, this is in fact a recursion in the coefficients.

While for any given $\ell$ the outer sum (using $n_1{+}...{+}n_d{+}k=\ell$)
in (\ref{compare_coeff}) is finite, 
the inner sum (using $i_1,...,i_d\in\Nat$) nevertheless usually forces us to do an infinite summation.

In the general case, trying to compute the coefficients this way would involve
non-algebraic methods: To compute the sum in equation (\ref{compare_coeff}) 
we need to show that the coefficients $c_{\nu,k,i_1,...,i_d}$ converge to $0$ 
quite fast, a similar situation to the summation of a power series. 
From the experience in \cite{Mue93}, we make the conjecture that this is true and that
a uniform polynomial time complexity of the coefficients should lead to polynomial 
time complexity of the sums $\sum c_{...}...$ as well as of the sequence $(a_\ell)$.
In this paper we do not want to generalise the lengthy proof for the one-dimensional systems 
given in \cite{Mue93},
but rather have a closer look at several special cases that lead to further efficiency.

\subsection{Zero vector as initial value}

If we not only use $t_0=0$ but additionally restrict the initial value $\overline{w_0}$ to $(0,\ldots,0)$, (\ref{compare_coeff})
can be simplified significantly.
In this special situation  we have $a^{(1)}_{\nu,0}=a_{\nu,0}=w_{\nu,0}=0$.
In the recursion (\ref{coeff-recursion}) this implies $a^{(2)}_{\nu,1}=a^{(2)}_{\nu,0}=0$, then 
further $a^{(3)}_{\nu,2}=a^{(3)}_{\nu,1}=a^{(3)}_{\nu,0}=0$ etc., by induction we get $a^{(i)}_{\nu,n}=0$
for $i>n$. This finally reduces (\ref{compare_coeff}) to the following form:
\begin{equation}
\label{compare_coeff1}
a_{\nu,\ell+1}=\frac{1}{\ell+1} 
\sum_{\substack{k,n_1,n_2,...,n_d\in\Nat\\ n_1+...+n_d+k=\ell}}~~
\sum_{0\leq i_1< n_1,...,0\leq i_d< n_d}
 \left[c_{\nu,k,i_1,...,i_d}\cdot a^{(i_1)}_{1,n_1}\cdot\ldots\cdot a^{(i_d)}_{d,n_d}\right] ~.
\end{equation}
Equations (\ref{coeff-recursion}) and (\ref{compare_coeff1}) together provide us with 
a finite recursion scheme to compute all coefficients of the Taylor series. 
Using exact real arithmetic like the iRRAM software this can be implemented quite straightforward.

\subsection{Non-zero initial time}
\label{n_z_i_t} Additionally to an arbitrary  $\overline{w_0}$ we might also consider 
arbitrary time instants $t_0$ for the initial condition  $\overline{w_0}=(w_{1,0},\ldots,w_{d,0})$
such that $y_\nu(t_0)=w_{\nu,0}$. Usually, this is ignored as we are able 
to transform the  flow functions $F_\nu$ to 
$$E_\nu(t,x_1,\ldots,x_d):= F_\nu(t+t_0,x_1+w_{1,0},\ldots,x_d+w_{d,0})$$
and consider the following system: 
\begin{equation}\label{IVP1} 
\dot z_\nu(t) = E_\nu(t,z_1(t),\ldots z_d(t))~~~~,~~~~ z_\nu(0)= 0 
~~~~~(1{\leq}\nu{\leq} d) ~.
\end{equation}
Let $z_\nu(t)$ be the  solution functions for this system.
The original system using the $F_\nu$ and  $y_\nu(t_0)=w_{\nu,0}$ is then solved by the functions $y_\nu(t)$ defined as 
$$ y_\nu(t) := z_\nu(t-t_0)+w_{\nu,0} ~.$$
Unfortunately, we now need the coefficients of power series for the $E_\nu$, that is for $F_\nu$
 centered at $(t_0,\overline {w_0})$ and not at the origin $(0,\ldots,0)$. 
How to get these coefficients depends on how $F_\nu$ is given.

\begin{itemize}
\item If $F_\nu$ is given via the series (centered at $(0,\ldots,0)$), 
we could do a re-expansion in the new center, which is essentially a composition of a power series and 
a linear function. Here \cite{vdH08} could be a good starting point, where the composition
of two power series is considered (but just for one variable).
\item If $F_\nu$ is given via an algorithm computing the function in a neighborhood of the initial value,
we could try to develop $F_\nu$ into a series, similar as in \cite{Mue93}.
\end{itemize}
In any case, trying to compute the coefficients for $E_\nu$ will force us to use 
non-algebraic methods and to evaluate quite complicated infinite sums. 
Our implementation is not yet mature enough to treat such complicated cases;
in the following we look at cases where we can have arbitrary initial values 
without the need to do a complicated transformation of the power series.

\subsection{Autonomous linear systems}

A very important class of applications are linear systems of differential equations, {\em i.e.},
each function $F_\nu$ is actually linear in the arguments $x_i$ :
 $$ F_\nu(t, x_1,\ldots x_d)=  f_{\nu,0}(t) + f_{\nu,1}(t)\cdot x_1+\ldots + f_{\nu,d}(t)\cdot x_d ~.$$
In this case, the IVP is already reduced to a system of $n^2+n$ functions $f_{\nu,i}:\Real\to\Real$.

Note carefully that $f_{\nu,i}$ do not need to be linear themselves.
They still can be very complex: in our setting they would be holomorphic 
(but now in just one variable $t$). These linear systems lead to a coefficient system with 
$$ c_{\nu,k,i_1,..,i_d}\neq 0 ~~~~\Longrightarrow ~~~~i_1+\ldots+i_d\leq 1 ~.$$
Still infinitely many coefficients could be non-zero (varying with $k$), leading to the necessity of 
an infinite summation  in (\ref{compare_coeff}).

In applications, a further property of the IVPs under consideration might be helpful: often the systems are 
autonomous, {\em i.e.}, the flow conditions do not depend on the time parameter $t$. This implies that each function $f_{\nu,i}$ actually has to be constant, so in this case  our coefficient system satisfies the following restriction:
\begin{equation}
\label{fully_linear} c_{\nu,k,i_1,..,i_d}\neq 0 ~~~~\Longrightarrow~~~~ i_1+\ldots+i_d\leq 1\wedge k=0 ~.
\end{equation}
In this case we only have to consider  $d^2+d$ numbers (instead of functions)
defining the IVP.

Together with the recursion for the $a^{(i)}_{\nu,n}$ from 
equation (\ref{coeff-recursion}),  we get the following finite(!) recursion scheme for the computation of all
coefficients of the solution of the IVP:

\begin{equation}
\label{lin+hom}
a_{\nu,\ell+1} = \frac{1}{\ell+1} 
\sum_{\substack{n_1,n_2,...,n_d\in\Nat\\ n_1+...+n_d=\ell}}~~
\sum_{\substack{i_1,i_2,...,i_d\in\{0,1\}\\ i_1+\ldots+i_d\leq 1}}
\left[c_{\nu,0,i_1,...,i_d}\cdot a^{(i_1)}_{1,n_1}\cdot\ldots\cdot a^{(i_d)}_{d,n_d}\right] ~.
\end{equation}
This formula can be further reduced using that the values  $a_{\nu,n}^{(0)}$ in
(\ref{coeff-recursion}) are very simple: As soon as a term in (\ref{lin+hom})
contains a component $a^{(i_j)}_{j,n_j}$ with $i_j=0$ and $n_j>0$, the product is zero.
On the other hand, the condition $i_1+\ldots+i_d\leq 1$ implies that at most 
one index $i_j$ can be non-zero. 

In consequence, for $l>0$ we get the recursion

\begin{equation}
\label{lin+hom2}
a_{\nu,\ell+1} = \frac{1}{\ell+1} 
\sum_{1\leq j \leq d}
\left[c_{\nu,0,\underbrace{\mbox{\scriptsize $0,0,...1$}}_j,0,...,0}\cdot a^{(1)}_{j,l}\right] ~.
\end{equation}
Additionally, for $l=0$, we have one further term:
\begin{equation}
\label{lin+hom3}
a_{\nu,1} = 
c_{\nu,0,0,...,0}
+
\sum_{1\leq j \leq d}
\left[c_{\nu,0,\underbrace{\mbox{\scriptsize $0,0,...1$}}_j,0,...,0}\cdot a^{(1)}_{j,0}\right] ~.
\end{equation}
Please note that for these autonomouse linear systems, the power series for the flow functions trivially 
have an infinite radius of convergence. Furthermore, the initial condition $(t_0,\overline{w_0})$ can now be
arbitrary: we may use the transformation from the previous section, but only applied to the time parameter $t$
(as we are able to treat non-zero $\overline{w_0}$ directly). But then, due to the autonomous system, 
the coefficients for the transformed system and the original system  are identical; we only have to solve 
$$\dot z_\nu(t) = F_\nu(t,z_1(t),\ldots z_d(t))~~~~,~~~~ z_\nu(0)= w_{\nu,0}
~~~~~(1{\leq}\nu{\leq} d) ~.
$$
using (\ref{initial condition},\ref{coeff-recursion},\ref{lin+hom2},\ref{lin+hom3}). 
This gives us a power series for each $z_\nu$ and we simply have use 
\begin{equation}\label{lin_transformation}
 y_\nu(t) := z_\nu(t-t_0) ~.
\end{equation}

\subsection{Nonlinear type, not autonomous, using multinomial flow functions}
From the previous section it is clear how to get a bigger class of IVPs that can be treated algebraically.
Suppose there is some $\mu\in\Nat$ such that $c_{\nu,k,i_1,\ldots,i_d}$ is zero as soon as one of its indices 
is larger than $\mu$.
Then the infinite sum $\sum_{i_1,\ldots,i_d\in\Nat}...$ in (\ref{compare_coeff})
reduces to just a finite sum $\sum_{i_1,\ldots,i_d\leq \mu}...$. Additionally, re-centering the coefficients
like in Subsection \ref{n_z_i_t} is just a finite manipulation of polynomials. 

In consequence, the current version of our implementation contains an IVP solver based on the following summary of the considerations in this section:
$$\fbox{\begin{minipage}{.95\textwidth} \begin{itemize}
\item Suppose in (\ref{IVP0}) only  a finite number of coefficients $c_{\nu,k,i_1,\ldots,i_d}$ are non-zero. 
\item 
Suppose that the initial value $(t_0,\overline{w_0})$   as well as the  $c_{\nu,k,i_1,\ldots,i_d}$  are uniformly computable. 
\item 
Then the power series $a_{\nu,n}$ for the solutions of the IVP are uniformly computable.
\end{itemize}
\end{minipage}}$$

\section{Taylor Sequences and Bounds}
The previous section has shown how we can get access to the coefficients of the power series for the IVP solution.
An important additional part of the evaluation of the IVPs is, of course, the necessary summation
of these sequences. Here we can obviously not avoid to compute sums of infinitely many values. 
The computational complexity of such a summation has been addressed for example in  \cite{Mue93}: if the coefficients of the 
series are uniformly computable in polynomial time, then the sum function also has polynomial complexity in the interior of the circle of convergence. To show this, a very detailed consideration of all intermediate rounding errors was necessary.
This would also be necessary in an implementation if we use a traditional multi-precision package for the computation.
Fortunately, an implementation using exact real arithmetic is much simpler in this regard, 
as the software package is able to
deal with all these cumbersome details by itself 
and we can instead concentrate on the more important issue: the truncation error
coming from using a finite summation instead of an infinite one.

So consider a sequence $(a_k)$ of Taylor coefficients together with a radius $R\in\Real$ 
such that $\sum a_k\cdot z^k$ 
converges absolutely  (in the complex plane) for any $z\in\Cplx$ with $|z|\leq R$ to a function $f$. 
Please note that essentially a simple linear transformation of the argument is sufficient 
if we have a series of the form $\sum a_k\cdot (z-z_0)^k$.

In finite time, we are only able to compute partial sums $\sum_{k=0}^n a_k\cdot z^k$ from the sequence, 
leading to truncation errors of $|\sum_{k=n+1}^\infty a_k\cdot z^k|$. However, we additionally just need a computable upper bound for this truncation error that converges to zero with increasing $n$ to implement
the infinite sum in exact real arithmetic. Suppose we have access to an upper bound $M\in\Real$ for $|f(z)|$ on $\{z\in\Cplx:|z|=R \}$.
Using the Cauchy integral formula, we see that 
$|a_n|\leq M\cdot R^{-n}$ holds for any $n$. This gives rise to the following explicit error
formula, valid in case of $|z|< R$:

\begin{equation}\label{truncation error}
\begin{array}{lcl} \displaystyle
|\sum_{k=n+1}^\infty a_k\cdot  z^k|
&\leq& \displaystyle \sum_{k=n+1}^\infty |a_k|\cdot|z^k|
~~~~~\leq~~~~~ \sum_{k=n+1}^\infty M\cdot R^{-k}\cdot|z^k|\\
&\leq& \displaystyle M\cdot\left(\frac{|z|}{R}\right)^{n+1}\cdot 
\sum_{k=0}^\infty\left (\frac{|z|}{R}\right)^{k}
~~~~~=~~~~~\frac{M\cdot R}{R-|z|} \cdot\left(\frac{|z|}{R}\right)^{n+1} ~.
\end{array}
\end{equation}
So, for an approximation with a truncation error of $\leq2^{p}$ we just 
have to add all the terms $a_n\cdot z^n$ until $\frac{M\cdot R}{R-|z|}\cdot\left(\frac{|z|}{R}\right)^{n+1}$ 
becomes smaller than $2^{p}$. 

The key to this summation is knowledge about  pairs $(R,M)$ 
with  $|f(z)|\leq M$ on $\{z\in\Cplx:|z|=R \}$.
In our case, the functions $f$  under consideration are solutions $y_\nu$ of IVPs; 
we can construct such bounds using the underlying flow conditions $F$, if we have access to a bound for $F$.

For any compact complex set $C\subset\Cplx^{d+1}$ 
the maximum $\mu(F,C):=\max_{\xi \in C,1\leq\nu\leq d}|F_\nu(\xi)|$ 
exists as soon as the $F_\nu$ are continuous on $C$; 
$\mu$ as a functional is even computable using standard representations for its arguments.
For  $(t_0,\overline w_0)$ and arbitrary $\delta,\epsilon>0$ we now 
consider the special complex neighborhood 
\begin{equation}\label{complex neighbor}
C:=C_\Cplx((t_0,\overline w_0),\delta,\epsilon):= \{(t,\overline z)\in\Cplx^{d+1}:|t-t_0|\leq 
\delta \wedge |\overline z - \overline{w_0} |\leq \epsilon\} ~.
\end{equation}
For the invariant domain $D=X$ we obviously have  $C\subseteq X$;
for smaller $D$ we would have to restrict $\delta$ and $\epsilon$ to 
$\delta,\epsilon\leq d_{X\setminus D}(t_0,\overline w_0)$ to ensure  $C\subseteq X$.

Unfortunately, using a general algorithm to compute $\mu(F,C)$ 
from its arguments $F$ and $C$ could be very time consuming. 
Furthermore, we obviously do not need the exact maximum but only a (good) upper bound.
So, for an efficient implementation,
it is important that the flow conditions $F$ are not only given with algorithms computing $F_\nu$ or
the coefficients $c_{\nu,k,i_1,\ldots,i_d}$, but also with an additional function $U_F$ within $\mu(F,C)\leq U_F((t_0,\overline w_0),\delta,\epsilon)$.

As long as a trajectory (considered as a function of a complex time variable!)
does not leave this set $C$, it cannot change faster than given by the bound $\mu(F,C)$.
So if we take the pair $(R,M)$ given by 
\begin{equation}\label{URM}
\begin{array}{lcl} \displaystyle
U&:=&U_F((t_0,\overline w_0),\delta,\epsilon) ~,\\
R&:=&\min\{\delta,\epsilon/U\} ~,\\
M&:=& |w_\nu|+R\cdot U ~,
\end{array}
\end{equation}
we have $|y_\nu(z)|\leq M$ for all $z\in\Cplx$ 
with $|z-t_0|\leq R$.

\section{Meeting the Guard}

With a combination of the two previous sections, we are able to compute the solution of many interesting IVPs 
 (at least on a small interval). Concerning the intended application to hybrid systems, we additionally want to 
find the point where a trajectory hits the guard for the first time.
Using the function $\Delta(t)=d_G(t,\overline y(t))$ of the distance between the guard set and the trajectory 
at time $t$, 
this is equivalent to finding the supremum $t_G:=\sup\{t\geq t_0\mid (\forall t',t_0\leq t'\leq t)\Delta (t')>0\}$. 

To find $t_G$, we restrict ourselves to computably closed guard sets $G$, 
so the distance function $d_G$ is a computable real function with  $G=d_G^{-1}(\{0\})$. 
In this case, $t_G$ is left computable \cite{Collins-2008a}.
In the following we present an algorithm to actually approximate $t_G$ from the left; 
later we also discuss an additional part
of the same algorithm that can sometimes  deliver approximations from the right, so that 
in the well-behaved case we even get a computable $t_G$.

\subsection{Steps}
Using the initial condition $\overline w_0$ at $t_0$ we want to find a $t_1$ such that $\Delta(t)>0$ for all $t\in(t_0,t_1)$. First we have to choose $\delta,\epsilon>0$ 
such that the (complex) neighborhood 
$C_\Cplx((t_0,\overline w_0),\delta,\epsilon)$ from (\ref{complex neighbor})
lies in $D_\Cplx$. 
Then we compute  $U$, $R$, and $M$ as given in (\ref{URM}); let $R'=R/2$.

Let  $t_1:=t_0+s_1$ for
\begin{equation}s_1:=\min\{\Delta(t_0)/U,R'\} ~.\label{stepsize}
\end{equation}
Using $R$ and $M$, the previous sections allow us to compute the value 
$\overline w_1:=\overline y(t_1)$ of the trajectory at $t_1$. 
Additionally we are sure that
$U$ bounds the flow functions in that region so that additionally between $t_0$ and  $t_1$ the trajectory may not touch 
the guard; {\em n.b.} using $R'=R/2$ above is not crucial, we only need to be sure that 
$s_1<R$  for (\ref{truncation error}).

We may continue this process, having two options: we may extend the solution obtained so far using the already known
Taylor series (algorithmically quite inexpensive, but only giving quite small extensions) or we may
determine new Taylor series (algorithmically expensive, but leading to bigger extensions). These two options will
be called `small steps' and `big steps'.
\begin{enumerate}
\item `small steps': If  $t_1$ is still sufficiently smaller than $t_0+R'$,
we may determine the distance $\Delta(t_1)$ between the guard and $(t_1,\overline w_1)$ 
to find a new step size $s_2= \min\{\Delta(t_1)/U,R'-s_1\}$ and let $t_2:=t_1+s_2$.
We can compute $\overline w_2:=\overline y(t_2)$ and may even 
iterate this process leading to sequences $(s_i)$  of step sizes and $(t_i)$ of  time instants with 
$t_i<t_G$.

Please note that here we are able to use the Taylor coefficients over and over again 
that have been computed from $(t_0,\overline w_0)$.  As always $t_{i}<t_{i+1}$, it might
be necessary to add further coefficients or to provide them with higher precision.
But as we approach the guard, the step width $t_{i+1}-t_i$ will converge to 0, so quite often we will need only a few (or even none) new additional coefficients. 
\item `big steps': As the resulting $t_i$ grow towards $t_G$, the summation 
of the series (centered in $t_0$) gets increasingly difficult. But of course, we are free
to use any $(t_i,\overline w_i)$ as a new initial condition.
This involves the necessity to compute new Taylor coefficients for  $\overline y$ 
at the new center $(t_i,\overline w_i)$.

The computation of these coefficients is quite time consuming, but on the other hand, afterwards 
the evaluation is faster again because of an improved truncation error (\ref{truncation error}).
Please note that now we will also have to recompute $U$, $R$, and $M$.

The optimal point for the re-centering of the problem surely depends on the flow conditions $F$ and
also on the guard $G$. Later we will briefly mention a corresponding heuristic.
\end{enumerate} 
Using this computation (in small or big steps) of points on the trajectory, we get an approximation of $t_G$ from the left.

\subsection{Traversing the guard}
When additionally trying to approximate $t_G$ from the right, we are in a situation 
similar to the computation of roots
of functions, as  $t_G=\min\{t\geq t_0\mid \Delta (t)=0\}$.
The comprehensive analysis of root finding in \cite{BHW07} helps to identify conditions we should use 
in order to allow a successful computation  of $t_G$. 
In general, we will not be able to check whether these conditions are met,
so we can only try to find right approximations to $t_G$.

The most helpful condition for root finding is that the underlying function should `change sign'. 
As the distance function $d_G$ (and hence also $\Delta$) we used so far is non-negative, we should have
further information on the guard set $G$:
If additionally the interior $G^o$ of the guard is computably open, the complement $X\setminus G^o$
is computably closed and the distance $d_{X\setminus G^o}$ is computable, too.
Then $\gamma_G(\xi):=d_G(\xi)-d_{X\setminus G^o}(\xi)$ is zero only at the border of $G$.
So instead of $\Delta(t)=d_G(t,\overline y(t))$ we should rather use $\Gamma(t):= \gamma_G(t,\overline y(t))$. 
This of course requires that the trajectory can be extended into the interior of $G$,
which is trivial if the invariant domain is $D=X$. Otherwise, the hybrid system must
be given accordingly.

If $\Gamma(t)$ really changes sign at $t_G$, the trajectory at $t_G$ is not a tangent to the border of $G$.
Unfortunately a corresponding test is uncomputable in general. At the moment we can only assume that 
the hybrid system and the initial condition are such that the sign change happens. If not, our algorithm will 
just compute the left approximation from the previous subsection.

On the other hand, if $\Gamma(t)$ changes sign at $t_G$, then
$\Gamma(t)>0$ for $t<t_G$ and there is an $\epsilon>0$ with $\Gamma(t)<0$ for $t\in(t_G,t_G+\epsilon)$.
Standard search methods can now be used to compute $t_G$; here we want to propose a more elaborate method:
A usual assumption in the world of {\small\verb#double precision#} arithmetic is that the distance between the trajectory and the guard is differentiable with a derivative significantly different from zero at $t_G$. 
In the following we translate this optimistic approach to the world of exact real arithmetic.

So suppose the guard has a smooth surface, such that the distance
to the guard is a continuously differentiable function (in $\Real^{d+1}$).
As the trajectory is differentiable, the 
function $\Delta(t)$, defined above as  the distance between the guard and the trajectory at time $t$,
will be differentiable, too. This allows us to
estimate the time the trajectory still needs until traversing the guard: 
If $t_{i-1}$ and $t_i$ 
are consecutive time instants where we evaluate the trajectory (outside of  $G$), then 
$\partial_i:=(\Delta(t_i)-\Delta(t_{i-1}))/(t_i-t_{i-1})$ is the derivative $\partial_i=\dot \Delta(\zeta)$ for some time point 
$\zeta$ between $t_{i-1}$ and $t_i$, due to the mean value theorem.
Hence we may estimate $t_G\approx t_i+\rho_i$ for $\rho_i:=\Delta(t_i)/\partial_i$, and 
$\hat t_i:=t_i+2\cdot \rho_i$ is a candidate where $\Gamma(\hat t_i)<0$ might be true. 
We only have to be careful when $\partial_i$ is (almost) zero, for which we 
implemented the following algorithm: If our goal is to find $t_G$ with an error of at most $2^{-n}$ ,
then we first check whether surely $\rho_i<2^{-n-1}$; only in this case we really compute $\rho_i$ and
$\Gamma(\hat t_i)$. So, together with strictly monotonic increasing left approximations $t_i$ for $t_G$, 
we can also construct a candidate list $(\hat t_j)$ of possible right approximations, all satisfying
$\hat t_j\leq t_j+2^{-n}$.

If we know for sure that $\Gamma(\hat t_j)<0$ for such a candidate $\hat t_j$, we know $t_j< t_G<\hat t_j$; 
so we have an approximation to $t_G$ with error $2^{-n}$ allowing us to terminate the algorithm.

Unfortunately, we cannot check easily whether $\Gamma(\hat t_j)<0$, as such a test is not computable. 
As a substitute for this un-feasible test we use multivalued tests 
whether $\Gamma(\hat t_\ell)<-2^{-k}$ or  $\Gamma(\hat t_\ell)>-2^{1-k}$ with a precision $k$. These tests
can be computed in finite time and they are applied as follows: 
whenever we compute a new pair $(t_j,\hat t_j)$ we check all previously computed 
candidates $\hat t_\ell$ (i.e.\ for $\ell<j$) again, 
but now with higher precision $k:=j$.
As $j$ goes to infinity, we will eventually find any $\hat t_l$ such that $\Gamma(\hat t_\ell)<0$; 
as soon as the first is found, our algorithm stops with success. 

The efficiency of these repeated tests is greatly improved, if we remove all those candidates $\hat t_j$ from the list where we found (in a similar way) that  $\Gamma(\hat t_j)>0$. 
Additionally, we may remove all candidates $\hat t_j$ from the list, 
as soon as a value $\rho_i$ is again larger than 
$2^{-n-1}$.

\section{Prototypical implementation}

We used the iRRAM package to implement a prototype for the proposed algorithm, whose core structure uses dynamically constructed functions of types
\verb#FUNCTION<int,vector<REAL> ># (for sequences of real vectors) and   \verb#FUNCTION<REAL,vector<REAL> ># (for vector-valued functions on the real numbers). 
The implementation of such function objects in an imperative language like \verb#C++# has been 
described in \cite{Mue09}, it is based on a lazy evaluation technique.
Thus we avoid the necessity to implement explicit (and computationally very expensive) representations 
for functions and sequences given in \cite{BHW07,We00}, {\em e.g.}.

Using corresponding constructors \begin{itemize}\item  \verb#a=ivp_solver_simple (w,F)# 
yielding a vector power series \verb#a# 
for flow conditions \verb#F# and an initial condition \verb#w# implementing equation (\ref{lin+hom}) ,
\item \verb#f=taylor_sum(a,R,M)# yielding the (vector-valued) 
sum function \verb#f# for a (vector-)power series \verb#a# and 
corresponding radius \verb#R# and bound \verb#M#,
and
\item \verb#w=f(bs)# evaluating \verb#f# at \verb#bs# with \verb#bs#$<$\verb#R# as a limit 
using equation (\ref{truncation error}) to control the truncation error,
\end{itemize}
the core of the implementation is essentially just the following loop of `big steps' interspersed with 
`small steps' as mentioned in the previous section:
{\small\begin{verbatim}
do { // big steps
  a = ivp_solver_simple (w,F);
    ... compute R,M ...
  f = taylor_sum(a,R,M);
  do { // small steps
      ... compute a step size from the distance to the guard ...
      ... accumulate the step size in a variable s  ...
      ... evaluate f(s) ...
      ... try whether a sufficiently good approximation has been found ...
      ... if yes: stop ...
  } until ( s is large enough for a big step ) 
  w= f(s) 
 }  
\end{verbatim}
}
As the evaluation of \verb#f(s)# is a core part here, it is important to get a reasonably efficient implementation
of the Taylor summation. The existing limit operators in the iRRAM package were not fitting, 
as they were not yet applicable to \verb#FUNCTION# objects 
(they were essentially only usable for predefined algorithms). 
Additionally, the general heuristic of the iRRAM (that tries to compute limits with the maximal used precision)
 lead to an enormous waste of time;
a new limit operator based on (\ref{truncation error}) had to be added. All other necessary operations 
were already present in the  published version of the package.

As a first benchmark we used the simple linear system 
\GLA
\dot y_1(t)= y_2(t)~~~~;~~~~ \dot y_2(t)= -y_1(t) + 0.02 \cdot y_2(t)~~~~\mbox{with} 
~~~~ t_0=0, \overline w_0=(0,1) ~.
\GLE
Without the term $c_{1001}=0.02$, the solution $\overline y$ would simply be the pair $(\sin,\cos)$;
with the additional term we still have an oscillation, but with a growing amplitude.

As guard set we chose $G=\{(t,x_1,x_2)\in \Real^3\mid x_1 \leq -2\}$. Here 
the question was simply to approximate the first $t_G$ where $y_1(t_G)=-2$ (we found $t_G\approx73.5422061995...$).
The size $s_1$ of the small steps was chosen as in (\ref{stepsize}); 
whenever the accumulated small steps 
grew larger than $\min\{\sqrt[4]{s_1}\cdot \sqrt{R},R'\}$, a big step was made.
This bound of the big steps was chosen heuristically as 
an attempt to match the much higher complexity of the IVP 
solution at big steps with the more frequent Taylor summations at small steps.

The following graph shows an 3d-plot of the resulting
trajectory constructed from  the (linearly interpolated) points $(t_i,\overline w_i)$.
\\[2ex]
%
%
%
%
\includegraphics[width=\textwidth]{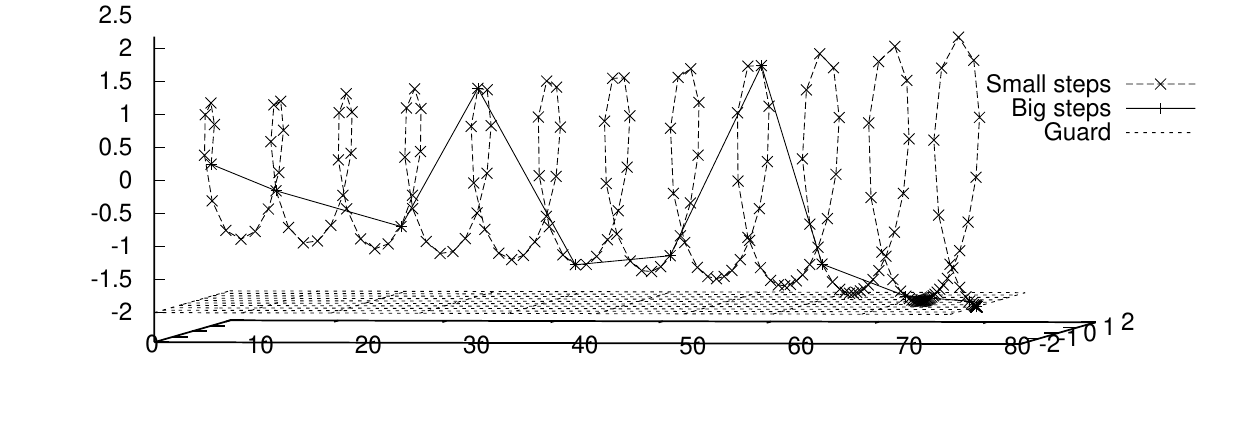}
\\[-1ex]
The following table shows a few results of computations with this IVP. Its interpretation is as follows: 
To approximate $t_G$ with an error of at most $2^{-n}$, the software chose a working precision of  $2^{-p}$, 
using $b$ big steps (re-evaluations of the IVP), $s$ small steps (evaluations of the Taylor sum) with a maximal
index of $\ell_{max}$ (working with an order of $\ell_{max}$) and took time $t$ (on an AMD Athlon 64X2 Dual Core Processor 4600+).

$$\begin{array}{|c||c|c|c|r|r|}\hline
\mbox{result bits } n&\mbox{working bits } p& \mbox{\#big steps } b& \mbox{ \#small steps } s &\ell_{max}&\mbox{time }t \\\hline
20   & 242  &  9 & 223  & 108  & 0.271s\\\hline 
50   & 242  & 10 & 283  & 108  & 0.298s\\\hline
100  & 242  & 10 & 384  & 108  & 0.297s\\\hline
1000 & 1332 & 12 & 2200 & 430  &  5.42s\\\hline
10000& 11787& 14 & 20361& 3506 & 308 s\\\hline
\end{array}$$
To compare our results we used the IVP solvers from the popular high-level language \verb#octave#,
that is primarily aiming at numerical computations ({\small \verb#www.gnu.org/software/octave#}), 
in order to solve the above IVP. 
We applied them just to approximate the trajectory starting from $t_0=0$ up to $73.543$.
Only between $73.542$ and $73.543$ we tried to find the point where it dropped below -2 
(without even trying to verify that this was the first solution).
Within a few milliseconds, a naive application of the solver gave 
a result near $t'_G=73.54225$. As only 6 decimals were in common with our result of $t_G=73.5422061995...$, 
we tried less naive ways, which initially produced the same result  $t'_G$.
Being convinced from the correctness of our own implementation, we continued playing with
the \verb#octave# solver; with further variations of its parameters applied in a quite elaborate way,
we were able to get results different from both $t_G$ and $t'_G$.
The best combination we found resulted in  $73.542208$, now 
with $7$ correct decimals, but within $0.7s$ of computation time. 
As the results from the \verb#octave# solver quite erratically jumped around $73.5422$ with further variations
of the parameters, we believe that more than 6 decimals precision cannot reliably be expected. 
Our conclusion from these experiments is that solving IVPs might be an area 
where exact real arithmetic can actually compete with ordinary {\small\verb#double precision#} arithmetic 
in terms of speed and precision.

To illustrate the effect of varying distances $|t_G-t_0|$ on our algorithm,  we removed the perturbating coefficient $0.02$.
Additional we chose the guard set $G_\eta=\{(t,x_1,x_2)\in \Real^3\mid t\geq \eta\}$ for a given $\eta$
and just printed 9 leading decimals of $y_1(t_{G_\eta})$. 
As  $t_{G_\eta}=\eta$, this setting transformed our algorithm into a slow (but still exact) 
method to compute $\sin(\eta)$. The results in the following table show that further reductions in the error propagation in our software are necessary before it can really be applied for larger ranges of $\eta$.
Again we compared our results with the \verb#octave# IVP solver, which was significantly faster for larger $\eta$ but had problems with its precision again.
$$\begin{array}{|c|r||c|c|c||c|r|}\hline
\eta&\multicolumn{1}{c||}{\sin(\eta)}&\multicolumn{3}{c||}{\mbox{our implementation}}&\multicolumn{2}{c|}{\mbox{\texttt octave}}
\\
&&\mbox{working bits } p&\mbox{\#big steps } b&\mbox{time } t& \mbox{time}&\multicolumn{1}{c|}{\mbox{result}}\\\hline
10       & -0.544021110 & 136  &  2  &  0.02s  & 0.007s & -0.5440211'86\\\hline
100      & -0.506365641 & 242  & 10  &  0.2s   & 0.06s  & -0.50636'2329\\\hline
1000     &  0.826879540 & 1737 & 95  &  17.5s  & 0.55s  &  0.8268'84089\\\hline
10000    & -0.305614388 & 14807& 681 &  3706s  & 5.3s   &  -0.305'931729\\\hline
\end{array}$$

\section{Summary}
In this paper we analysed a recursive method for ODE solving with an emphasis on algorithmic applicability.
As the method consists of two basic parts, the construction of Taylor series and their subsequent summation,
we were able to adopt it to special requirements found in hybrid systems.

Based on our benchmarks we conjecture that a closer analysis of the complexity of the algorithm  will show that for given poly-time computable $F$  and initial values the computed value $t_G$ has complexity polynomial in the precision, if the prerequisites of the algorithm are given. An open question is, 
whether this very specific aspect of precision-oriented complexity can also be expressed in a uniform 
way depending on $F$ or at least on the initial
value. It might be much easier to construct a dependency on the actual value of the derivative of $\Delta$ at $t_G$.
Additionally, the effect of the heuristic for the relation between the `small steps' and the `big steps' on the complexity is also worth studying.

Apart from these questions of computational complexity, further detailed considerations for the cases 
of non-linear equations and esp.\ for the non-algebraic solutions are important, as well as for the
influence of the dimension of the state space onto the efficiency of the algorithm.

Considering the trajectories within a single component of the state space is obviously only an initial step 
into hybrid systems. The far goal is to improve the efficiency of reachability analyses: for trajectories starting in a given subset of states we want to know all reachable states, including those induced by the jumps between the different 
components of the  state space. Of course, treating single trajectories like we did it in this paper can be extended to sets, as computability implies effective continuity.
Our last benchmark concerning the solution of an IVP on longer 
time intervals however shows that the modulus of continuity derivable from our algorithm does not yet allow 
an efficient set-oriented evaluation. In the near future, we want to consider the influence of Lipschitz properties in the dependency between the initial condition and the final state at $t_G$ to improve this. 
Additionally, we will address efficient data structures for closed sets together with corresponding set-valued functions.

\end{document}